# THERMODYNAMIC ANOMALIES AND STRUCTURAL FLUCTUATIONS IN AQUEOUS SOLUTIONS OF TERTIARY BUTYL ALCOHOL

Deepa Subramanian[1,2], Jeffery B. Klauda[1], Jan Leys[2,3], and Mikhail A. Anisimov[1,2]

[1]*Department of Chemical and Biomolecular Engineering, University of Maryland, College Park, MD 20742, USA*

[2]*Institute for Physical Science and Technology, University of Maryland, College Park, MD 20742, USA*

[3]*Laboratorium voor Akoestiek en Thermische Fysica, Departement Natuurkunde en Sterrenkunde, KU Leuven, Leuven, Belgium*

**Abstract**

In this work, we discuss the connection between the anomalies of the thermodynamic properties, experimentally observed in tertiary butyl alcohol (TBA) – water solutions, and the molecular clustering in these solutions, as revealed by molecular dynamics (MD) simulations. These anomalies are observed in relatively dilute solutions of about 0.03 – 0.08 mole fraction of TBA and become more pronounced at low temperatures. MD simulations show that these solutions exhibit short-ranged (order of 1 nm), short-lived (tens of picoseconds) "micelle-like" structural fluctuations in the same concentration range. We attribute the anomalies in the thermodynamic properties of aqueous TBA solutions to these structural fluctuations on the molecular scale.

## 1. Introduction

Aqueous solutions of alcohols are of great interest in various fields, ranging from biotechnology and metabolic engineering to materials science. The physical chemistry of aqueous solutions of alcohols has been studied for a long time [1,2]. Low-molecular-weight monohydric alcohols such as methanol, ethanol, propanol, isopropanol, and tertiary butyl alcohol are completely miscible with water [1]. However, these solutions show anomalous behavior in their thermodynamic properties, such as a minimum in excess volume, a maximum in heat capacity, or a negative excess chemical potential of water [1]. All these thermodynamic properties display a strong dependence on the solute concentration. Amongst such monohydric alcohols, the most pronounced anomalies are observed in aqueous tertiary butyl alcohol solutions. Hence in this work, we focus on the thermodynamic properties of tertiary butyl alcohol – water solutions and relate these properties to structural fluctuations on molecular scale.

Tertiary butyl alcohol (TBA) or 2-methylpropan-2-ol is a special alcohol. The hydroxyl group of the TBA molecule, which promotes dissolution in water, counteracts the hydrophobic *tert*-butyl group of the molecule [3]. TBA is the highest molecular weight alcohol to be completely miscible with water in all proportions under ambient conditions [4]. Aqueous TBA solutions show anomalies in many thermodynamic properties, namely, excess molar and excess partial molar volumes, excess molar and excess partial molar enthalpies, excess molar and excess partial molar entropies, excess molar and excess partial molar Gibbs energies (excess chemical potentials), activity coefficients, heat capacity, isothermal compressibility, and volume expansivity [1,2,5]. All these properties show a strong dependence on solute concentration, with contrasting behavior in the solute-poor and solute-rich regions. The experimental results are often interpreted in terms of a common clustering in aqueous TBA solutions [6-10], a concept supported by computer simulations [11-17]. In this work, we present new results from molecular dynamics simulations of aqueous TBA solutions carried out with the TIP4P/ICE water model.

## 2. Anomalous Thermodynamics of TBA – Water Solutions

*2.1 Phase diagram of TBA-water solutions*

Various research groups have investigated the phase diagram of the TBA - water system over the years [4,18-20]. The liquid-solid phase diagram at ambient pressure, as determined by Kasrian *et al.* by differential scanning calorimetry, is shown in Figure 1 [4]. Liquid TBA and water are completely miscible with each other under ambient conditions. The system exhibits two eutectics, one at about 0.057 mole fraction TBA in water at about −5 °C and another one at about 0.68 mole fraction TBA in water at about −3 °C. The phase diagram of TBA-water was also investigated at higher pressures, up to 200 MPa, by Woznyj *et al.* [20]. Raising the pressure shifts the eutectic points to a lower temperature but at about the same concentration.

*2.2 Excess volumes*

Excess properties and excess partial molar properties provide valuable information on solute-solvent interactions within a solution. In ideal solutions, with no preferential interactions between the components of the mixture, the excess volume of mixing is zero. However, for associated liquids, such as hydrogen-bonded solvent systems, significant negative excess volumes are observed [1,2].

The excess molar volume of TBA-water solutions, as shown in Figure 2(a), is a strong function of solute concentration, but weakly dependent on temperature [21]. The excess partial molar volumes of TBA and water, in TBA-water solutions as a function of TBA concentration, are shown in Figure 2(b) [21]. As seen from Figure 2(b), as the concentration of TBA in the solution is increased, the partial molar volume of TBA initially decreases, passes through a minimum in the water-rich region, and then increases. The minimum in the excess partial molar volume of TBA becomes sharper and more enhanced as the temperature is lowered [21]. The excess partial molar volume of water exhibits a maximum in the water-rich region and a minimum in the TBA-rich region. This minimum becomes more pronounced and shifts to higher TBA compositions as the temperature is increased [21]. We have checked the data in Figure 2(b) for thermodynamic consistency with the help of the Gibbs-Duhem relation. The data mutually agree within 10% accuracy.

A comparison of the partial molar volumes of TBA in aqueous TBA solutions with the partial molar volumes of other C1-C4 alcohols (such as methanol, ethanol, propanol, isopropanol, *sec*-butanol, ethylene glycol, glycerol) in their respective aqueous solutions shows that as the alcohol chain length and branching increases, the minimum in partial molar volume becomes sharper and more pronounced and shifts to a lower alcohol concentration [22]. The most significant effect is seen in aqueous TBA solutions [1,22].

*2.3 Excess enthalpies and Gibbs energies*

The excess enthalpy of mixing and the excess partial molar enthalpies of mixing are shown in Figure 3 [23-25]. The excess enthalpy of mixing is negative in the water-rich region, goes through a minimum at a TBA concentration of about 0.06 mole fraction and is then positive in the TBA-rich region with a maximum at about 0.75 mole fraction TBA. The excess partial molar enthalpy of TBA is a sharp function of concentration in the water-rich region, and becomes nearly independent of concentration at higher concentrations.

The excess molar Gibbs energy of mixing and the excess partial molar Gibbs energies of mixing (also known as excess chemical potentials) are shown in Figure 4 [23,26]. The excess molar Gibbs energy of mixing is positive and shows no significant anomalous behavior. However, the excess partial molar Gibbs energy of TBA shows a peculiar dip at about 0.05 mole fraction TBA. In the same range of TBA concentrations the chemical potential of water shows a negative deviation from ideality at lower temperature as is discussed in the following section.

*2.4 Activity coefficient of water at low temperature*

Activity coefficients are uniquely related to the excess chemical potentials as $\mu_i^E = RT \ln \gamma_i$, where $\mu_i^E$ and $\gamma_i$ are the excess chemical potential and activity coefficient of species $i$ in solution, $T$ is the temperature, and $R$ is the universal gas constant. Experimentally determined activity coefficients of water in TBA-water solutions at 10 °C are shown in Figure 5 [5,27]. The logarithm of the activity coefficient of water in the water-rich region of the mixture shows a pronounced negative deviation from ideality, with a minimum at about 0.04 mole fraction TBA, and rapidly increases towards positive deviations at higher TBA concentrations. A similar trend is also seen in other aqueous C1-C4 alcohol solutions [27].

However, thermodynamic consistency with the data for the excess partial molar Gibbs energies at 25 °C, discussed in the previous section, suggests that the minimum of the logarithm of the activity coefficient of water should disappear at higher temperatures.

*2.5 Heat capacity and isothermal compressibility*

Heat capacity and isothermal compressibility measurements yield important mixture information about solution behavior as they are highly sensitive to structural changes in the solution. The isobaric heat capacity $C_P$ and the isothermal compressibility $\kappa_T$ are related as $C_P - C_V = VT\frac{\alpha_P^2}{\kappa_T}$, where $C_V$ is the isochoric heat capacity, $V$ is the volume, $T$ is the temperature, and $\alpha_P$ is the isobaric thermal expansivity [28]. The heat capacity and isothermal compressibility of TBA-water solutions as a function of TBA concentration are shown in Figures 6 and 7, respectively [29,30]. From Figure 6 it is seen that the isobaric heat capacity of the solution exhibits a sharp maximum in the water-rich region and rapidly decreases to the molar isobaric heat capacity of TBA as the concentration of TBA increases. As the temperature decreases the magnitude of this heat-capacity maximum increases. It is seen from Figure 7 that the isothermal compressibility exhibits a minimum at about 0.03 mole fraction TBA. Since $\alpha_P$ does not have a pronounced anomalous behavior [30], the maximum in the isothermal compressibility reflects the minimum in the heat capacity.

*2.6 Nature of the thermodynamic anomalies in aqueous TBA solutions*

A possible interpretation for the thermodynamic anomalies observed in aqueous alcohol solutions can be obtained by relating them to the hydrogen bond interactions between the alcohol and water molecules [1]. In the water-rich region of the solution, strong hydrogen bonds form between alcohol and water. This may lead to an initial volume contraction, as seen in Figure 2(a). Thus, in the water-rich region of the solution, structural stabilization becomes enhanced with increasing alcohol concentration, until a further increase of the alcohol concentration imposes a strain on the water network and its structural integrity breaks down.

The peculiar behavior of the activity coefficient of water is in agreement with the explanation given above. In the water-rich region, strong solute-solvent interactions in the form of strong hydrogen bonds between TBA and water are favored, leading to negative values for the logarithm of the activity coefficient.

Another explanation for the thermodynamic anomalies can be obtained via hydrophobic hydration [1,31,32]. Water molecules arrange themselves around a non-polar solute in a cage-like manner, with the solute molecules occupying the cavities of water. The negative values of the excess volumes (as seen in Figure 2(a)) can be attributed to the occupation of water cavities by alcohol molecules [1,22,31,32]. Thus the minima in excess molar and partial molar properties could be attributed to a "stoichiometric" (one TBA molecule per 15-17 water molecules) stabilization of the less-polar solutes by water [2,5].

The maxima in heat capacity observed at about 0.05-0.08 mole fraction TBA, as seen in Figure 6, were interpreted by the authors, as due to the presence of a structural transformation from an ordered state, at low TBA concentrations, to a disordered state at high TBA concentrations [29]. De Visser *et al.* attribute this behavior to solutes such as TBA, which enhance water structure at low solute concentrations [33]. Franks and Ives refer to these effects as being due to hydrophobic hydration [1]. In fact, such a trend in the heat capacity appears to be typical for hydrogen-bonding fluids [1,2,5].

The minimum of the compressibility of aqueous TBA solutions on the addition of small amounts of TBA may be attributed to the formation of a compression-resistant hydrogen-bond network water structure due to the making of hydrogen bonds with the solute [30].

## 3. Molecular Dynamics Simulations

*3.1 Methods*

In order to comprehend the origin of the thermodynamic anomalies in aqueous TBA solutions, molecular dynamics (MD) simulations on pure TBA and on aqueous TBA solutions have been carried out. It has been speculated that the mechanism for the anomalies in aqueous TBA solutions may have a similar origin as the formation of clathrate-hydrates [5]. This motivated the choice of the TIP4P/ICE water model used for these simulations as it accurately predicts the normal freezing point of water and of solid hydrate phases in water [34,35].

The NAMD simulation program [36] was used to perform all MD simulations with 2 femtosecond time steps. All TBA-water simulations consisted of 2188 molecules with mole fraction of TBA of 0.0380 and 0.0715. These simulations ranged from 200-1100 ns to probe the possibility of hydrate-like crystals. A single simulation of pure TBA at 283 K was also simulated with 192 molecules for 10 ns. The van der Waals interactions were smoothly switched off between 8 and 10 Å by a potential-based switching function. Long-range electrostatic interactions were calculated using the particle-mesh Ewald (PME) method [37]. An interpolation order of 4 and a direct space tolerance of $10^{-6}$ were used for the PME method. Parameters for TBA were taken from the CHARMM General Force Field [38]. Langevin dynamics was used to maintain constant temperatures for each system, while the Nosé-Hoover Langevin-piston algorithm [39,40] was used to maintain constant pressure at 1 bar. The Visual Molecular Dynamics (VMD) [41] program was used to create snapshots and to calculate the radial distribution functions (RDF).

*3.2 Results*

The radial distribution function (RDF) between the central carbons of TBA in pure TBA is shown in Figure 8(a). This figure shows the presence of a peak at a central carbon distance of 4.7 Å and a second peak at a distance of ~6 Å. The first peak in the RDF corresponds to strong van der Waals interactions, while the second peak corresponds to weak van der Waals interactions between the methyl groups of TBA. The RDFs between the oxygen atoms of TBA in pure TBA is shown in Figure 8(b). The initial large peak corresponds to strong hydrogen bonding between TBA molecules in pure TBA, but disappears when in aqueous solution. The simulation results from our work and from that of others [12-17] indicate that pure TBA exhibits a mixture of hydrogen-bonding interactions between the hydroxyl groups of the TBA molecules and non-polar interactions between the methyl groups of the TBA molecules.

In aqueous solutions, the RDFs between the carbon atoms in TBA, as seen from Figure 8(a), show that the first peak, which corresponds to van der Waals interactions in pure TBA, disappears when in aqueous solution and the second peak at a distance of about 6 Å becomes more prominent. Additional peaks are also calculated and differ from the pure TBA simulation. These simulations show that in aqueous solutions of TBA, relatively more TBA molecules surround a central TBA molecule as compared to the situation in pure TBA.

Bowron *et al*. have shown that in aqueous TBA solutions, the magnitude of the peak seen at ~6 Å gets enhanced as the concentration of TBA increases, as seen in Figure 9 [15]. This figure also shows that the magnitude of the second peak at ~8 Å decreases on increasing TBA concentration. The same behavior is observed in our MD simulations of aqueous TBA solutions.

The RDFs between the oxygen atoms of aqueous TBA are shown in Figure 8(b). As stated earlier, the first peak in pure TBA corresponds to strong hydrogen bonds between TBA molecules, which disappear when in aqueous solutions. Thus it is seen that in pure TBA hydrogen bonds are formed between the hydroxyl groups of the TBA molecules, while in aqueous solutions hydrogen bonds are formed between the hydroxyl groups of TBA and the water molecules. This correlates well with the behavior of the activity coefficient of water, discussed in Section 2.4.

The RDF between the oxygen atom on TBA and that on water is shown in Figure 10. There exists a strong peak at ~2.6 Å representing the strong hydrogen bonding of water to TBA. This replaces that of TBA-self hydrogen bonding and the closer distance than TBA-TBA hydrogen bonding suggest that water forms a stronger hydrogen bond with TBA. The higher order peaks (2-4$^{th}$ order) correspond to certain water structures around TBA. These structures are stronger at lower TBA concentration (0.038 mole fraction TBA) as seen from their higher peak in Figure 10.

*3.3 Interpretation of molecular dynamics simulations*

Snapshots from MD simulations in aqueous TBA solutions, shown in Figure 11, facilitate the interpretation of the RDFs. Figure 11 shows a snapshot from molecular simulations, indicating a dimer of TBA with van der Waals interactions between its methyl groups, surrounded by a hydrogen-bonded polygonal (either pentagonal or hexagonal) network between TBA-water and water-water molecules. The main RDF peak at ~6 Å, as seen in Figures 8 and 9, corresponds to the distance between central carbon atoms of TBA, which may be dimers, trimers or tetramers of TBA. The secondary peak at ~8 Å in Figures 8 and 9 corresponds to a distance between the central carbon atoms in oligomerized TBA with its nearest neighbor of unstructured TBA. The water molecules form a structure around TBA molecules, with the hydroxyl group of the TBA molecules forming one of the vertices of a hydrogen-bonded polygon.

The structural significance of the peaks calculated in the O(TBA)-O(Water) RDFs can also be visualized in snapshots of Figure 11. The primary peak in Figure 10 is from ~3 waters coordinating the hydroxyl group of TBA. The secondary RDF peak at ~4.4 Å and the tertiary peak at ~5.6 Å in Figure 10 correspond to larger distances between the vertices in hexagon and pentagon ring of water. These secondary and the tertiary peaks correspond to an additional ~17 – 21 surrounding water molecules. This leads to an effective "micelle" radius of 7.5 Å with the first water shell at 10.5 Å including the surrounding shells of water. Although the water structures surrounding TBA form polygons commonly found in hydrates [42], these were short-lived and appear to be transient, with an estimated life-time of the order of 10 – 50 ps.

Bowron *et al*. explain the effect of TBA concentration associated with molecular clustering [15]. They show that at low TBA concentrations (about 0.06 mole fraction TBA) solute-solute interactions are obtained through methyl group contacts, evidencing a significant association between TBA molecules. This association is spontaneous and persists for tens of picoseconds, as also seen from our MD simulations. This association occurs until the intermolecular solute-solute interactions can no longer be accommodated through the non-polar methyl-group interactions. As the TBA concentration is further increased, beyond 0.16 mole fraction TBA, the interactions between the TBA molecules resemble those in pure TBA.

**4. Discussion**

In this work, we connect the thermodynamic anomalies to the structural fluctuations in aqueous TBA solutions. As seen from MD simulations, at low TBA concentrations, aqueous TBA solutions show the presence of dynamic short-ranged "micelle-like" clusters, with a lifetime of about tens of picoseconds. These clusters tend to disappear as the TBA concentration increases. All these observations lead us to believe that at low TBA concentrations, TBA-water hydrogen bonds are preferred over TBA-TBA and water-water hydrogen bonds, leading to favorable solute-solvent interactions. As a result thermodynamic properties such as excess molar and excess partial molar volumes show minima and the excess enthalpy becomes negative in this region. At higher TBA concentrations, no such preferential TBA-water hydrogen bonding exists.

A more quantitative approach to explain this phenomenon can be obtained by examining the anomalies of the excess partial molar enthalpies. The minimum in the excess partial molar enthalpy of TBA is about $-17$ kJ/mol, as it is seen at about 0.06 mole fraction TBA (Figure 3(b)). This can be attributed to the hydrogen-bond energy between TBA and water molecules, which is stronger than the hydrogen-bond energy between water molecules, which is estimated to be about $-6$ kJ/mol [43,44].

Furthermore, the thermodynamic and structural anomalies of aqueous TBA solutions may significantly change the character of solubilization of hydrophobic compounds. Recently, it has been shown that the addition of small traces of a hydrophobe in aqueous TBA solutions, at low TBA concentrations (between 0.03 to 0.08 mole fraction TBA), leads to the emergence of inhomogeneities on the mesoscale [45,46]. In these ternary solutions, by using dynamic light scattering technique, Subramanian *et al.* have shown that in addition to molecular diffusion (with a decay time of about a few microseconds at the light-scattering wave number of $7 \cdot 10^6$ m$^{-1}$), there exists a slow relaxation process (with a decay time of a few milliseconds at the same wave number) corresponding to the diffusion of Brownian particles with a size of about 100 nm [45]. It has been further shown that these inhomogeneities are exceptionally long-lived and remain stable up to many months [47]. We speculate that the mesoscale inhomogeneities, observed in water-rich TBA solutions upon addition of hydrophobic compounds are akin to a microemulsion, with the hydrophobic molecules at the center of the droplet, surrounded by a hydrogen-bonded TBA-water network. Most recent measurements of isobaric heat capacity in TBA-water solutions confirm the earlier results [29] and show that the removal of the mesoscale inhomogeneities by repeated filtrations at low temperature does not significantly affect the anomaly in the isobaric heat capacity [48]. This reaffirms that the anomalies in the thermodynamic properties of aqueous TBA solutions are indeed attributed to a genuine microscopic structure (clustering) of the solution and not to accidental hydrophobic impurities that maybe present in commercial samples. However, this clustering significantly affects the way how hydrophobic molecules are solubilized in the solution. It is also worthy to note that the structural fluctuations at the molecular scale cannot be detected by dynamic light scattering technique because they are too fast (tens of picoseconds); however they may be possibly detected by some other special technique, such as neutron spin echo.

## 5. Conclusions

We attribute the thermodynamic anomalies seen in aqueous TBA solutions to structural fluctuations existing at the molecular scale. These structural fluctuations are different from the conventional fluctuations of concentration which are detectable by dynamic light scattering [45,46]. Favorable hydrogen-bonding interactions between TBA and water in the dilute TBA concentration regime (between 0.03 to 0.08 mole fraction TBA) lead to the formation of a short-ranged, short-lived "micelle-like" structure, where oligomers of TBA are surrounded by a hydrogen-bonded TBA-water network. This structure disappears as the TBA concentration increases because at higher TBA concentrations there are not enough water molecules available to form a network of similar structure. As a result, the thermodynamic anomalies exist in dilute TBA solutions. The nature of the "stoichiometric" TBA/water molecular ratio of 1/15-1/17, for which the anomalies are observed, is most intriguing. A possible explanation is to attribute it to the proposed two-state structure of cold and supercooled water [49]. Liquid water at low temperatures is viewed as a non-ideal "athermal solution" of two hydrogen-bond networks with different entropies and densities. The model predicts that upon the temperature increase the fraction of the low-density structure in water rapidly decreases. The lower-density, higher-entropy structure above the melting temperature contains only about 10-15 % of all water molecules [49]. This structure is full of cavities and might be mainly responsible for bonding and clustering with the TBA molecules.

Addition of a third compound could stabilize the structural fluctuations, leading to the formation of long-lived, stable mesoscale inhomogeneities, with a size of about 100 nm. Although the qualitative explanation of these phenomena, as presented in this work, sounds reasonable, a detailed quantitative description and modeling still need to be developed.

**Corresponding Author**

* Email: anisimov@umd.edu. Tel.: + 1-301-405-8049. Fax: +1-301-314-9404.

**Dedication and acknowledgments**

Mikhail Anisimov acknowledges long-term interactions and friendship with Professors Alexey Morachevsky and Natalia Smirnova. Their fundamental contributions in the area of thermodynamics of fluids and fluid mixtures have been a source of inspiration for this work.

The research of Deepa Subramanian and Mikhail Anisimov is supported by the Division of Chemistry of the National Science Foundation (grant no. CHE-1012052), USA. Jan Leys acknowledges the Research Foundation-Flanders (FWO) for a travel grant. Jeffery Klauda acknowledges computational time from the National Science Foundation through XSEDE resources provided by National Institute for Computational Sciences (Kraken) under grant number TG-MCB100139 and the High Performance Computing Cluster at the University of Maryland. We would also like to thank Elia Altabet and Sedef Ayalp for help with plotting some of the figures.

**Figure captions**

Figure 1:

Melting temperatures of TBA-water solutions as a function of TBA mole fraction. The horizontal and vertical lines divide the solid phase in regions where different ice and hydrate structures are assigned [4]. The data are reproduced after Ref. 4.

Figure 2:

(a) Excess molar volume of TBA-water solutions as a function of TBA mole fraction at 25 °C. (b) Excess partial molar volumes as a function of TBA mole fraction at 25 °C. Closed circles: TBA; open circles: $H_2O$. Both data sets are reproduced after Ref. 21.

Figure 3:

(a) Excess molar enthalpy of TBA-water solutions as a function of TBA mole fraction at 25 °C. (b) Excess partial molar enthalpies as a function of TBA mole fraction at 25 °C. Closed circles: TBA; open circles: $H_2O$. Both data sets are reproduced after Ref. 26.

Figure 4:

(a) Excess molar Gibbs energy of TBA-water solutions as a function of TBA mole fraction at 25 °C. (b) Excess partial molar Gibbs energies or chemical potentials as a function of TBA mole fraction at 25 °C. Closed circles: TBA; open circles: $H_2O$. The solid curve is the extrapolation of the regular behavior. Both data sets are reproduced after Ref. 26.

Figure 5:

Logarithm of the activity coefficients for $H_2O$ in TBA-water solutions as a function of TBA mole fraction at 10 °C. The dashed horizontal line indicates a value of 1 for the activity coefficient, corresponding to ideal mixing. The solid curves are guides to the eye. The data are reproduced after Ref. 27.

Figure 6:

Heat capacities of TBA-water mixtures as a function of TBA mole fraction at several temperatures. The curves are reproduced after Ref. 29.

Figure 7:

Isothermal compressibility of TBA-water solutions as function of TBA mole fraction at 25 °C. The data are reproduced after Ref. 30.

Figure 8:

Simulated radial distribution functions for pure TBA and two TBA-water solutions. (a) RDF between the central carbon atoms in TBA. (b) RDF between the oxygen atoms in TBA. Solid lines: pure TBA; dashed lines: 0.0380 mole fraction TBA in water; dotted lines: 0.0715 mole fraction TBA in water. The dashed and dotted curves have been shifted vertically for better visibility.

Figure 9:

Radial distribution functions between the central carbon atoms of TBA for different concentrations, simulated on the basis of experiment neutron scattering data [15]. Line: 0.06 mole fraction TBA in water; crosses: 0.11 mole fraction TBA in water; circles: 0.16 mole fraction TBA in water. Functions are offset vertically for clarity. Reprinted with permission from Ref. 15. Copyright 1998 American Chemical Society.

Figure 10:

Simulated radial distribution functions between the oxygen atom of TBA and the oxygen atom of $H_2O$ in TBA-water mixtures. Dashed line: 0.0380 mole fraction TBA in water; dotted line: 0.0715 mole fraction TBA in water. The dashed curve has been shifted vertically for better visibility.

Figure 11:

Snapshots from molecular dynamics simulations of 0.0380 mole fraction TBA in TIP4ICE water at 285 K, taken after 1000 ns. Distances are indicated in Å. (a) Distances between the central carbons in TBA and between TBA oxygens and close $H_2O$ oxygens indicated. (b) Distances between TBA oxygens and more distant $H_2O$ oxygens indicated.

**Figures**

Figure 1:

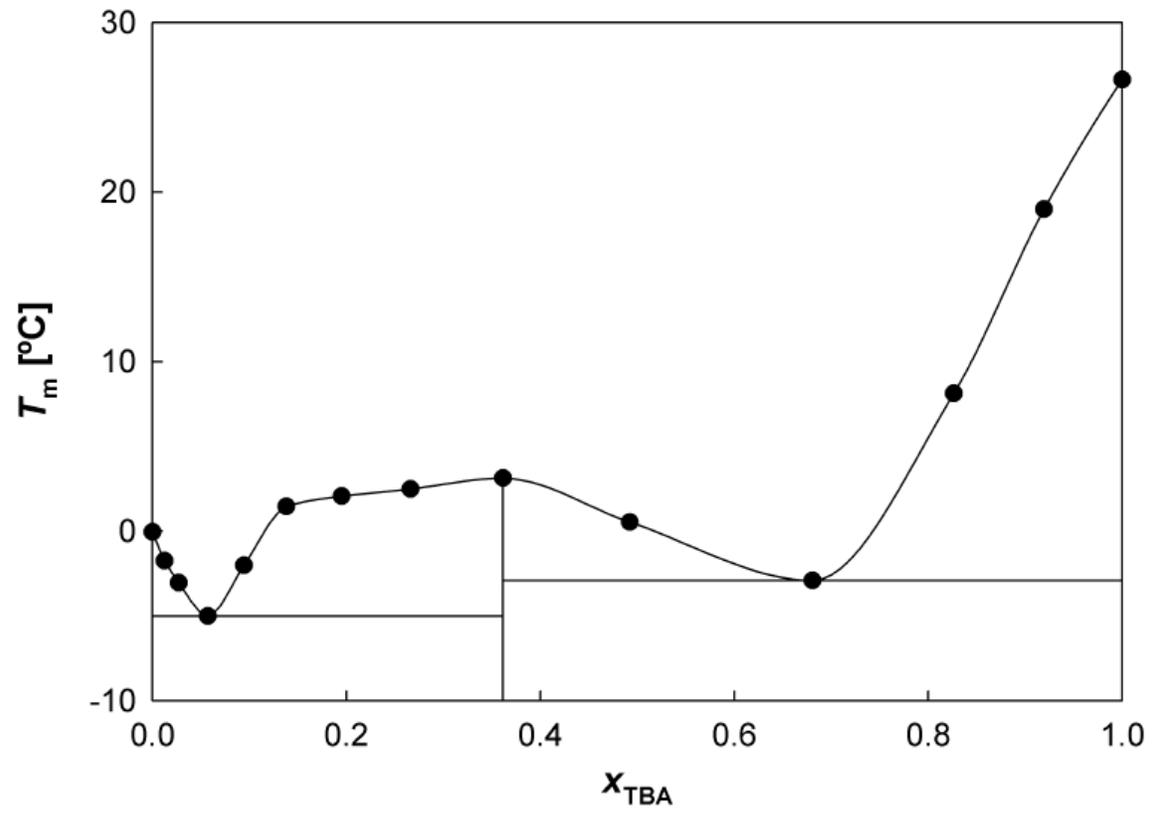

Figure 2:

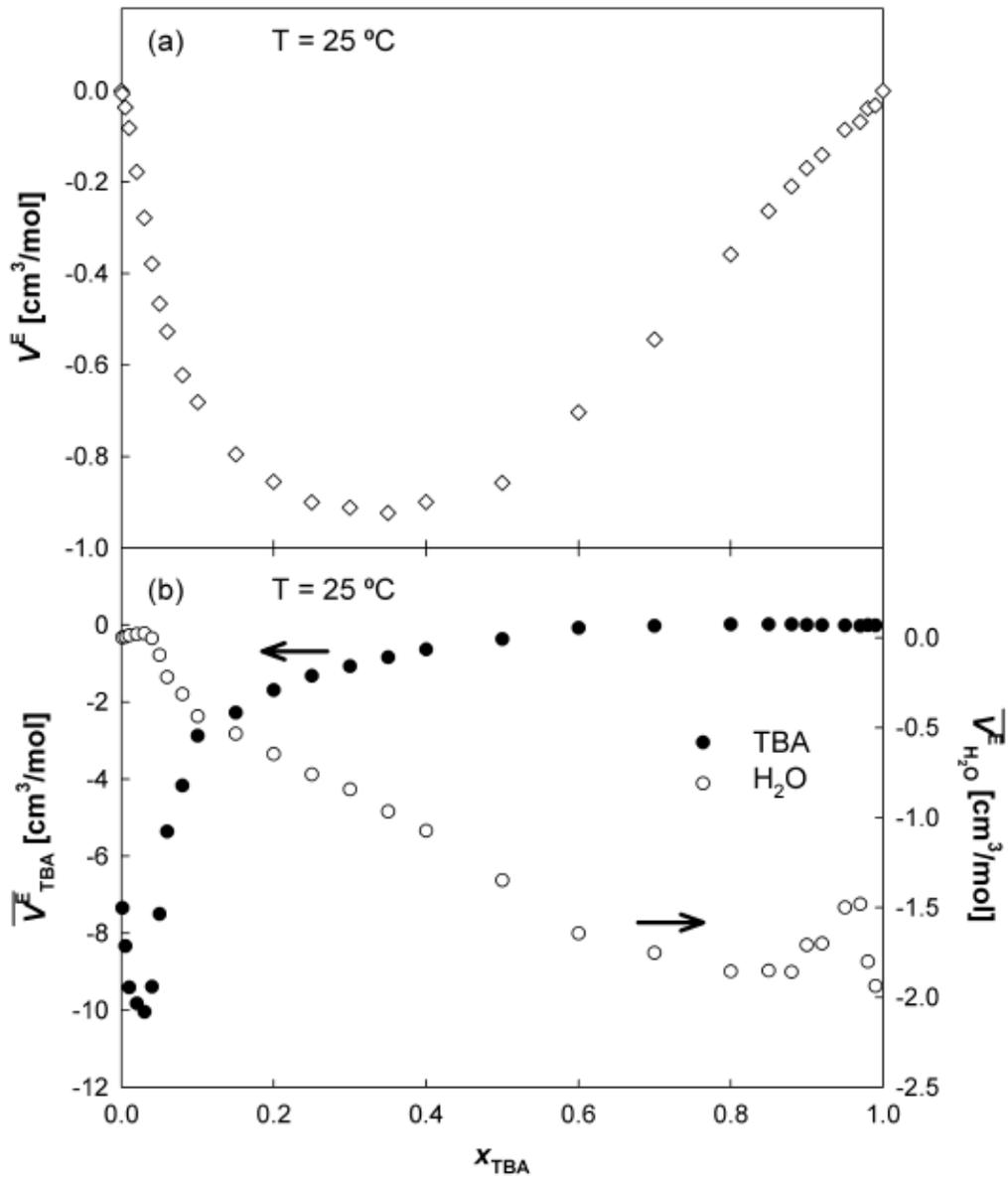

Figure 3:

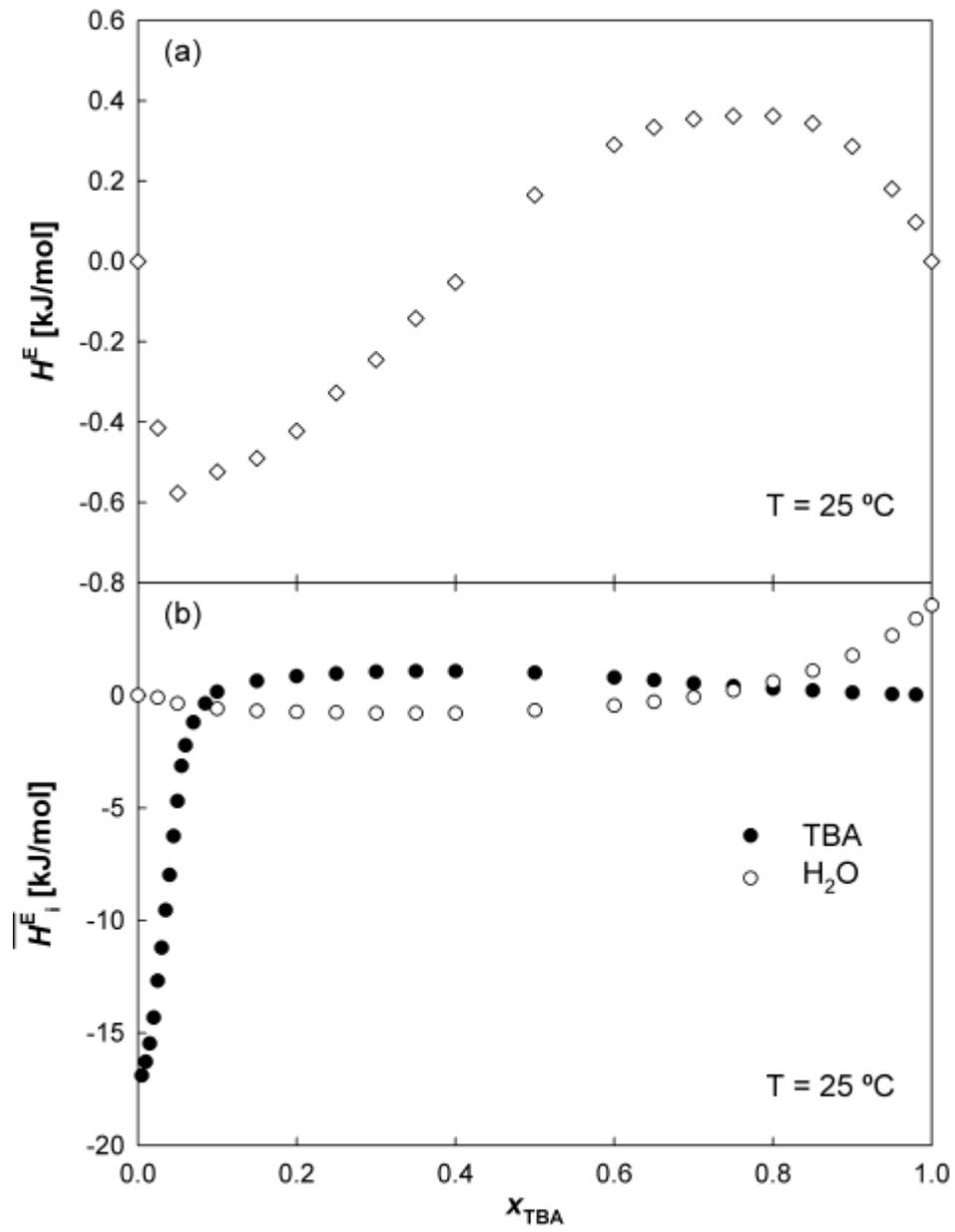

Figure 4:

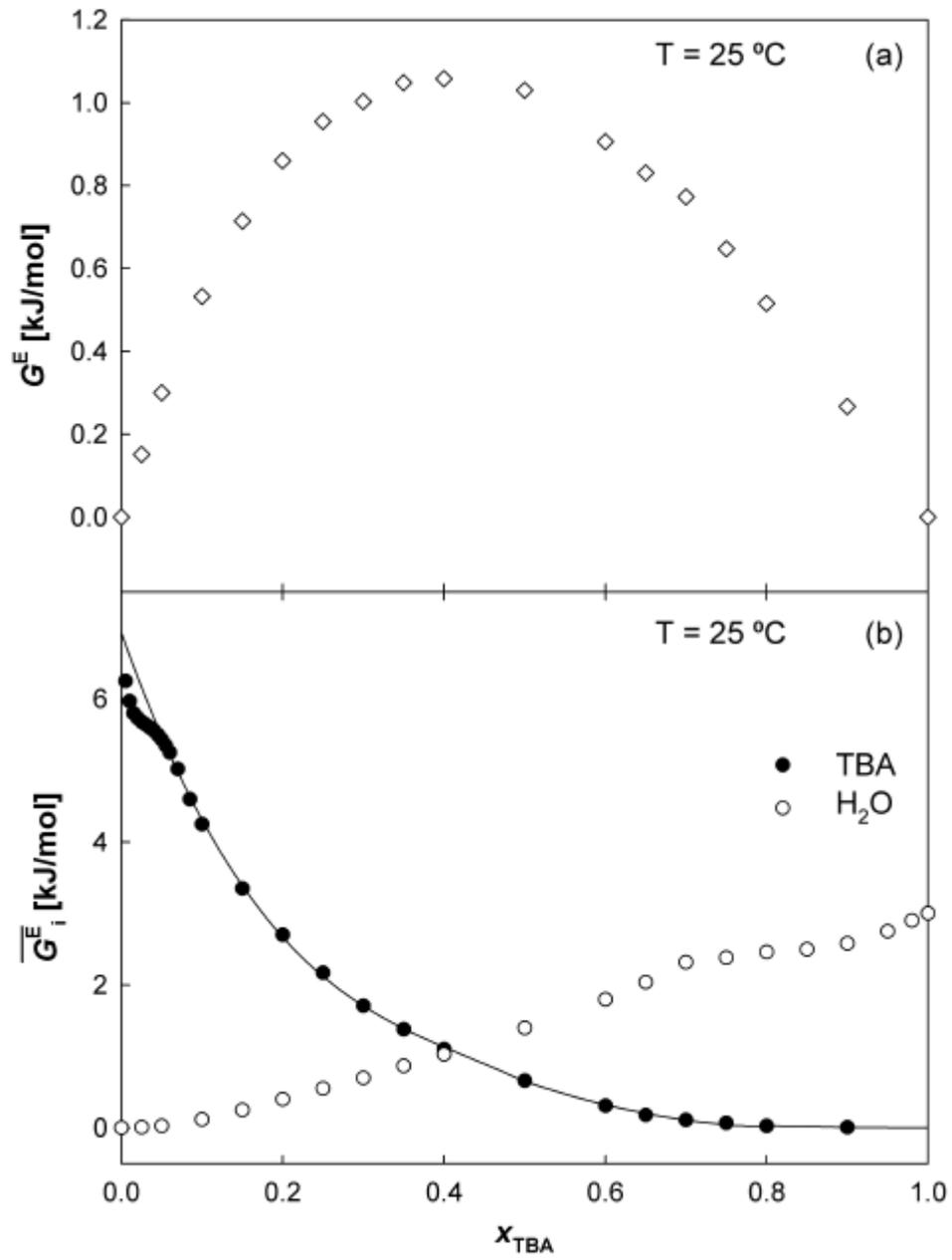

Figure 5:

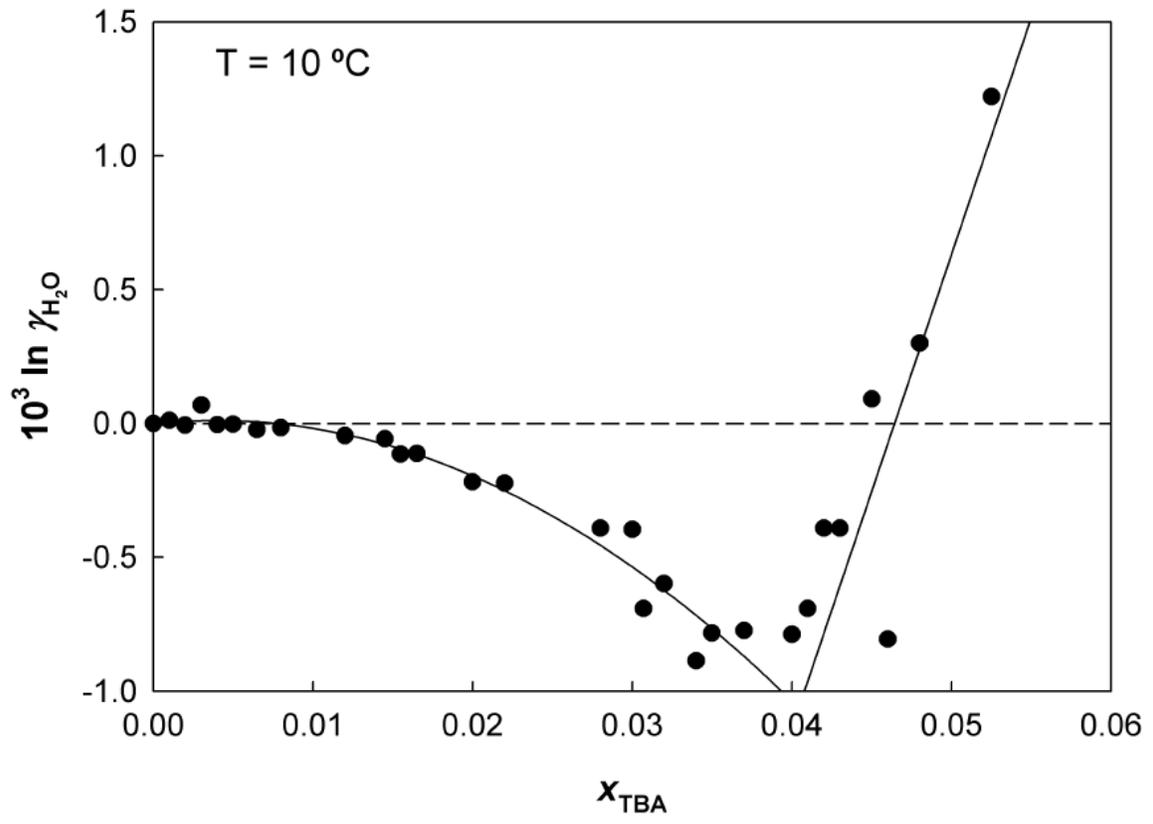

Figure 6:

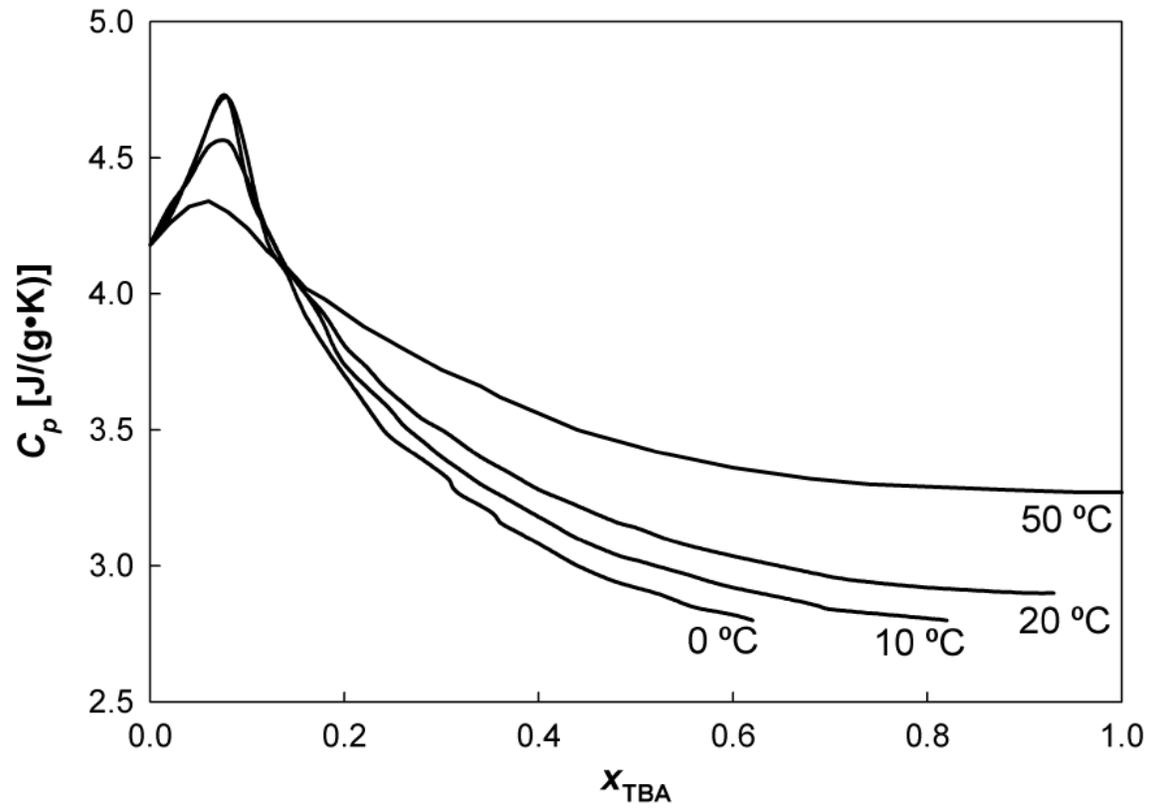

Figure 7

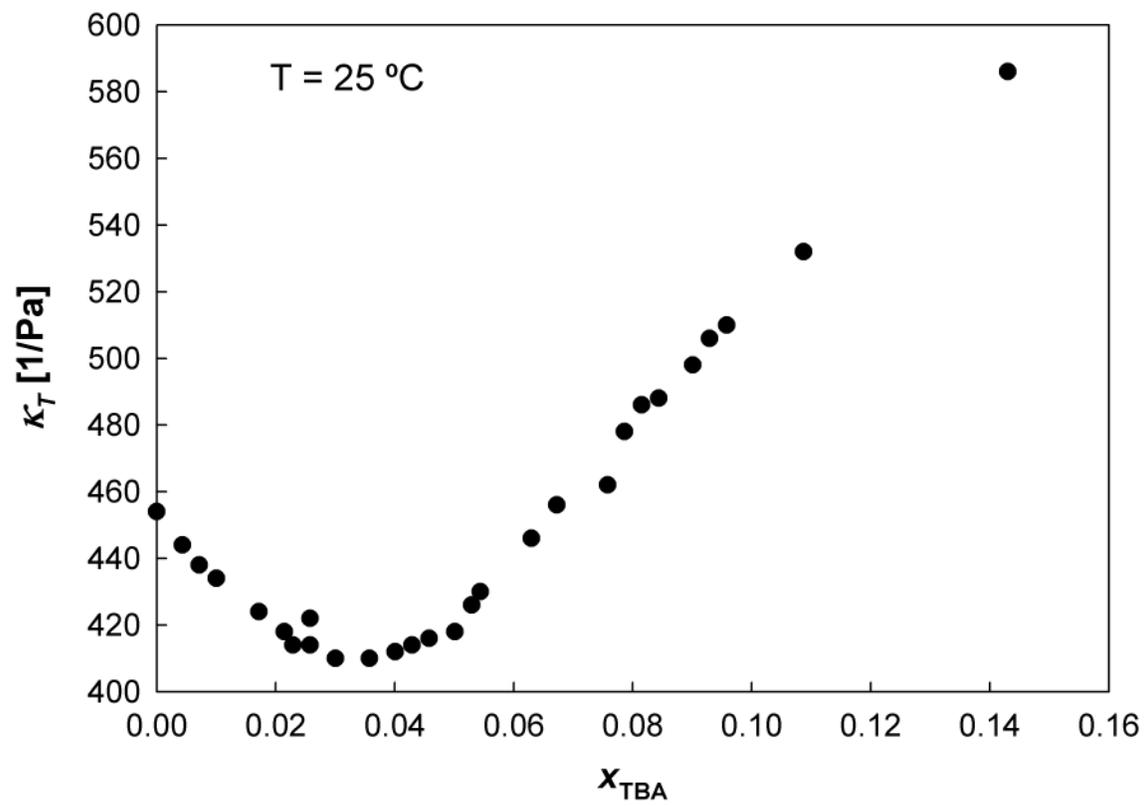

Figure 8:

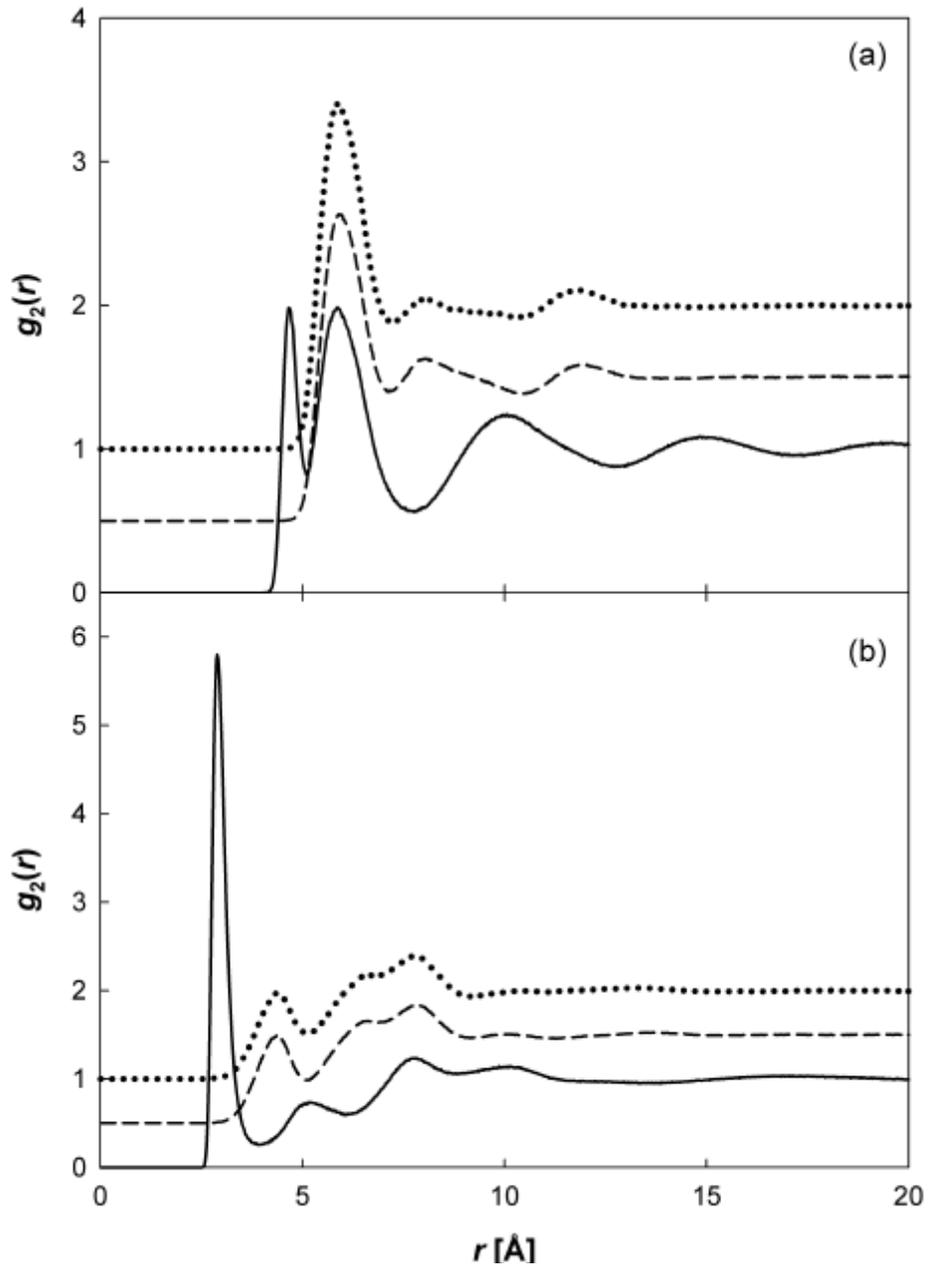

Figure 9:

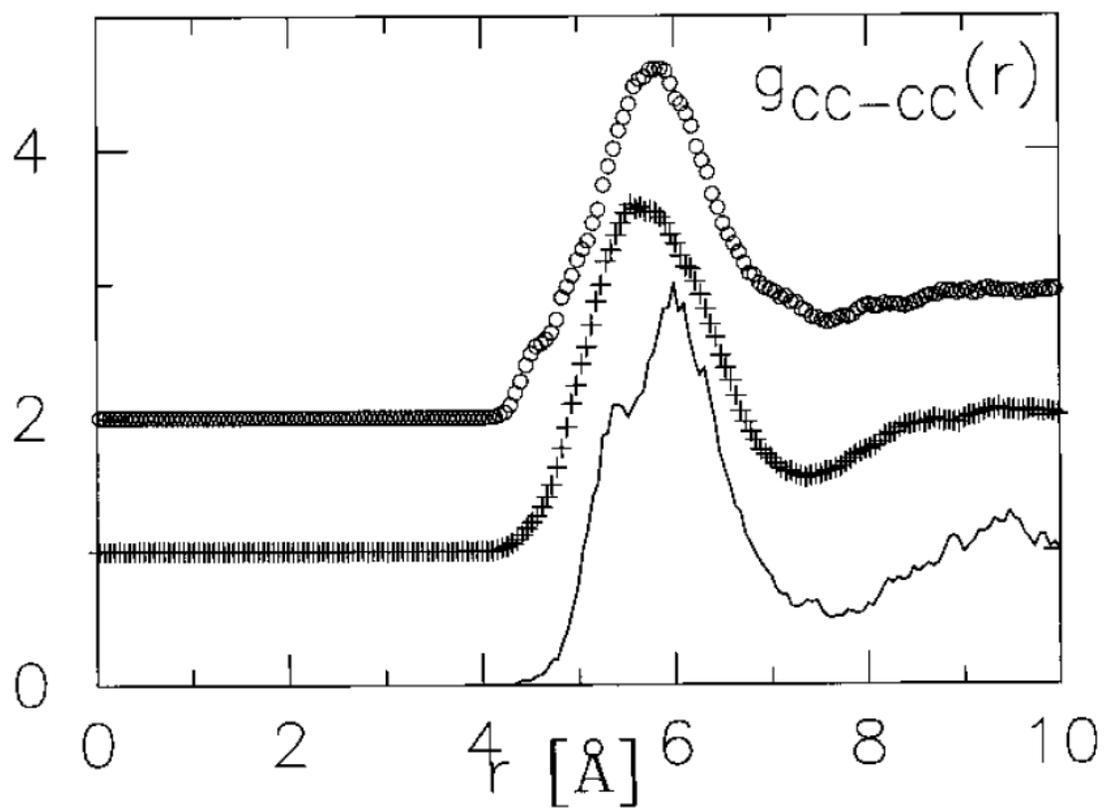

Figure 10:

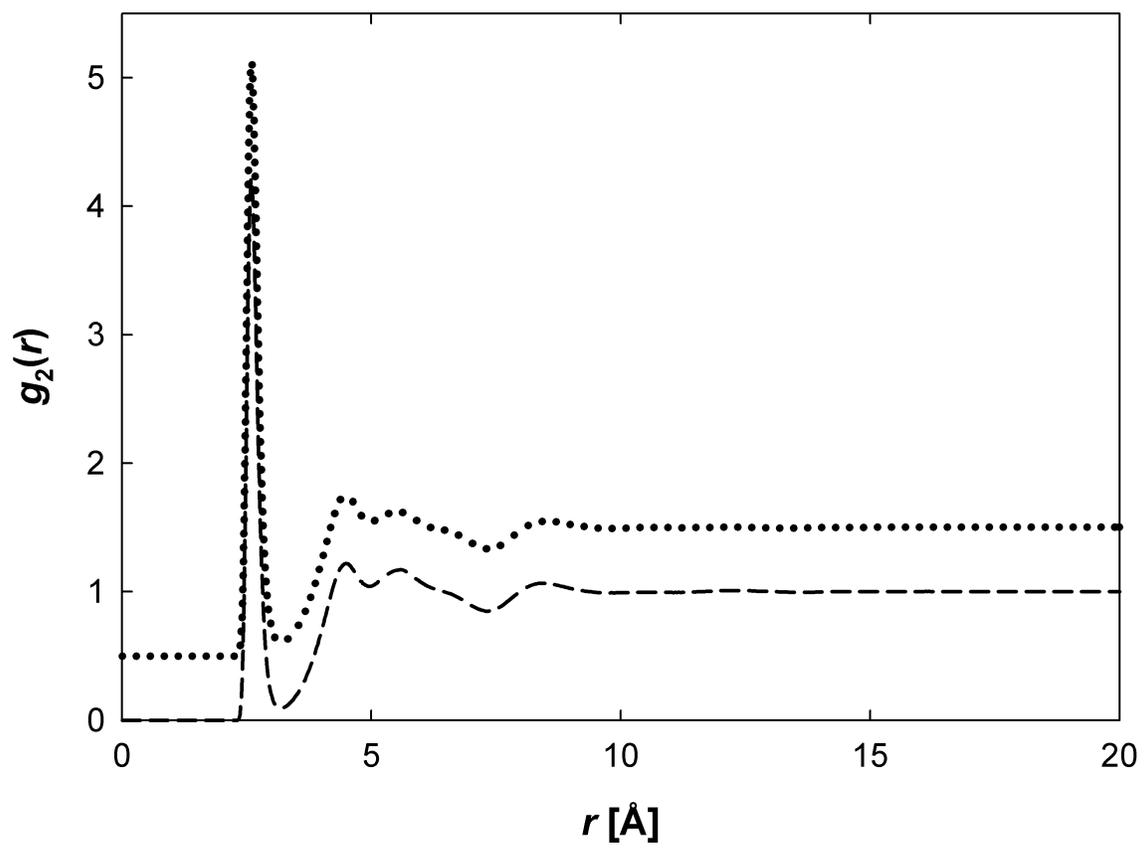

Figure 11:

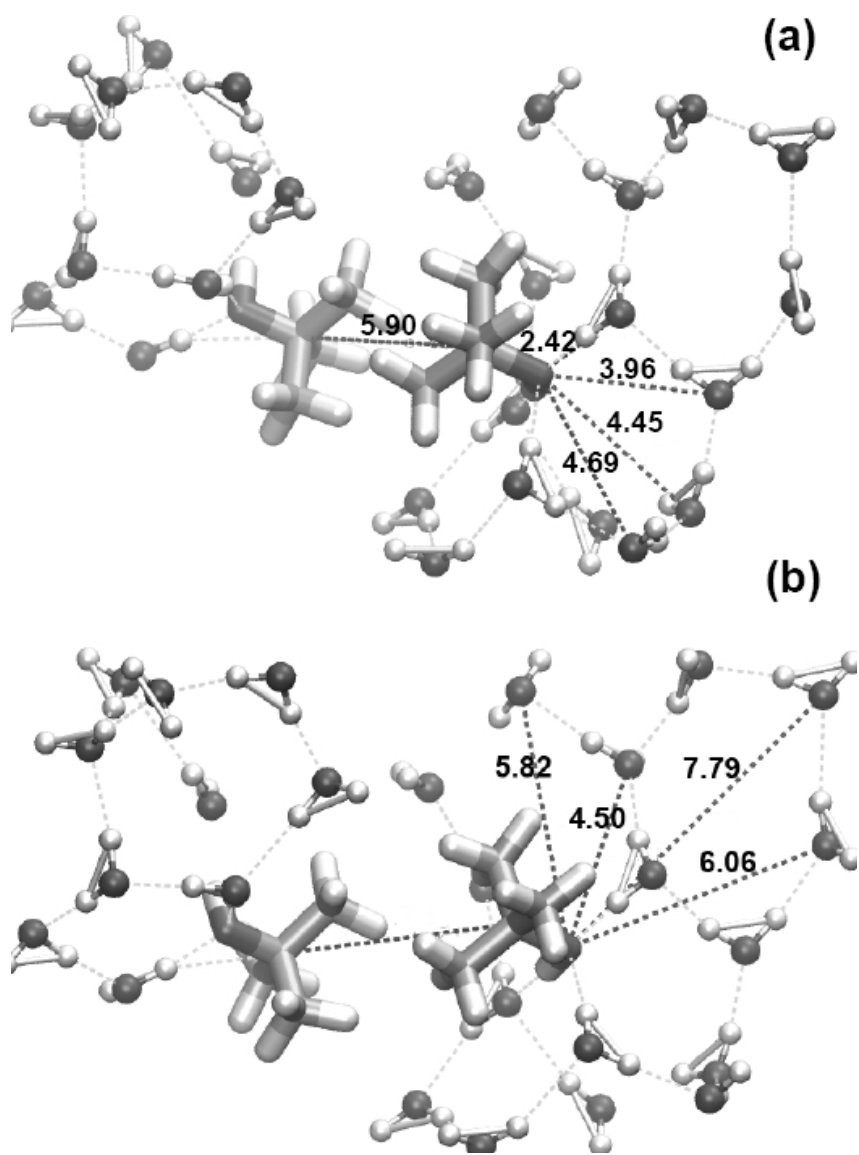